\documentclass[conference]{IEEEtran}
\IEEEoverridecommandlockouts
\usepackage{cite}
\usepackage{amsmath,amssymb,amsfonts}
\usepackage{algorithmic}
\usepackage{graphicx}
\usepackage{textcomp}
\usepackage{xcolor}
\def\BibTeX{{\rm B\kern-.05em{\sc i\kern-.025em b}\kern-.08em
    T\kern-.1667em\lower.7ex\hbox{E}\kern-.125emX}}
\begin{document}

\title{Text-Based Product Matching - Semi-Supervised Clustering Approach\\

}

%\author{\IEEEauthorblockN{Anonymous Authors \\ \\ \\ \\ \\}}

\author{
    \IEEEauthorblockN{
        Alicja Martinek\IEEEauthorrefmark{1}\IEEEauthorrefmark{2},
        Szymon \L{}ukasik\IEEEauthorrefmark{1}\IEEEauthorrefmark{2},
        Amir H. Gandomi\IEEEauthorrefmark{3}
    }
    \IEEEauthorblockA{\IEEEauthorrefmark{1} NASK National Research Institute, Poland}
    \IEEEauthorblockA{\IEEEauthorrefmark{2} AGH University of Krak\'{o}w, Poland}
    \IEEEauthorblockA{\IEEEauthorrefmark{3} University of Technology Sydney, Australia}
    Email: alicja.martinek@nask.pl, szymon.lukasik@nask.pl, amirhossein.gandomi@uts.edu.au
}

\maketitle

\begin{abstract}
Matching identical products present in multiple product feeds constitutes a crucial element of many tasks of e-commerce, such as comparing product offerings, dynamic price optimization, and selecting the assortment personalized for the client. It corresponds to the well-known machine learning task of entity matching, with its own specificity, like omnipresent unstructured data or inaccurate and inconsistent product descriptions. This paper aims to present a new philosophy to product matching utilizing a semi-supervised clustering approach. We study the properties of this method by experimenting with the IDEC algorithm on the real-world dataset using predominantly textual features and fuzzy string matching, with more standard approaches as a point of reference. Encouraging results show that unsupervised matching, enriched with a small annotated sample of product links, could be a possible alternative to the dominant supervised strategy, requiring extensive manual data labeling. 
\end{abstract}


\section{Introduction}
Online shopping and purchasing services via e-commerce platforms are constantly gaining popularity. It happens due to the expanding digitalization of the retail sector and the growing utilization of advanced algorithms embedded in such platforms. %This increasing trend was amplified by the global pandemic when clients started to value contactless transactions more. Interestingly economists do not expect to see the reverse effect or the plateau in the post-pandemic future \cite{tradegov:2022}.
According to Eurostat average share of European Union-based companies making a profit from e-commerce sales grew in ten years from 14\% to 19\% in 2021, peaking at 20.6\% in 2022 \cite{Eurostat:2022}. Same source reports a 20 \% growth in proportion of e-shoppers among all Internet users, from 55\% in 2012 to 75\% in 2022 \cite{Eurostat:2023grow}.
Meanwhile, in the United States, the share of online sales in the total market rose from 5.1\% to 15.4\% between Q1 2012 and Q2 2023 \cite{statista:2022}.
In such a setting, retailers are competing with each other not only in terms of profits, but also in cutting-edge technologies which drive the numbers.

The aforementioned increase in the number of transactions generates high volumes of data. This is where artificial intelligence thrives the most. So far, machine learning -- the common branch of AI -- has been used in many different areas of e-commerce. The most popular include, but are not limited to: recommendation engines \cite{rengines}, creating targeted marketing campaigns \cite{camp,adverti}, purchase predictions \cite{purchase}, dynamic pricing \cite{dynamicpricing2005,dynamic} and optimization of retailer resources \cite{reso}. A data-driven approach to e-commerce is beneficial for both sides of the deal. In the end, the customer receives a personalized and seamless experience whilst the retailer becomes more competitive amongst the other sellers.

Product matching, an instance of the common entity matching task, is one of the key exercises for retailers. Its goal is to match identical products from two different product feeds. This is not a trivial task given the nature of textual data. Product descriptions often require specialized background knowledge and can be characterized by several different modes. This all accounts for the importance of high-reliability product matching. Once completed it allows for comparisons between offerings of the same product, which can be further used for dynamic price optimization and selecting a fitted assortment for the client. Mentioned actions are means to fulfilling the ultimate goal of every company - profit maximization.

The aim of this paper is to show how product matching can be achieved with the semi-supervised clustering algorithm. Such an approach allows to exploit all benefits of the unsupervised methods. Keeping in mind that data labeling is expensive and time-consuming, it is our motivation to fully gain from widely available data and enrich it with a smaller sample of annotated data. It has been demonstrated that the described method increases the accuracy of generated clusters \cite{fixmatch:2020}. Our framework also includes text-mining algorithms used for feature engineering. We focus on similarity measures as they are go-to calculations for text comparison tasks.
The main contribution of this paper is to present a new view on the product matching problem. We do not invent a new algorithm but try to show fresh approach to handling such a task.

The paper is structured as below. Having given the introduction, Section 2 is a comprehensive review of current solutions to the product matching task. In addition, it also describes the feature engineering process. The following Section presents a deep dive into the proposed framework, delivering the details of the algorithm implementation. Preliminary results of conducted experiments are provided and thoroughly analyzed in Section 4. Finally, Section 5 covers conclusions and perspectives for possible future improvements to the proposed solution.

\section{Related Work}
\label{rw}
\subsection{Product matching}
Among the problems that can be solved with machine learning algorithms, product matching task is constantly \newpage
gaining importance due to exploding amount of data from online platforms. Retailers can leverage this information to better suit their offerings.

The core of product matching lies in obtaining pairs of matching goods, based on so-called product feeds. The feeds can originate from different sources, hence discrepancies in available attributes can become an obstacle. Such differences can manifest in distinct product taxonomies, reporting prices in varied currencies, including/excluding taxes or shipping costs, or inconsistent formulation of product names. The trick lies in conscious feature engineering that takes into consideration all possible data issues, not to mention missing information.

Carefully processed data that describes pairwise relations between items of merged feeds can be used to redefine the nature of the product matching problem. At this stage, the described task becomes a binary classification exercise. The target variable in such a setting takes a value of 1 when paired products represent the same physical good, and it becomes 0 otherwise. Models used for classification tasks usually output the probability of a record belonging to the given class (in this study, meaning that the products describe the same item).

There are many approaches to solving the assignment of entity matching. Most existing solutions are based on supervised learning methods. They include and are not limited to adapting XGBoost \cite{lukasik2021text}, using advanced natural language processing models as BERT \cite{tracz2020bert,peeters2020intermediate,Peeters_2022, ditto}, and incorporating deep neural networks \cite{li2020deep,ara,contrast} or fuzzy matching \cite{fuzzy:2014}. There even exist attempts to incorporate Large Language Models as chatGPT in the task of entity matching \cite{peeters2023entity, peeters2023using}.

Another important group of methods takes advantage of various types of data available in on-line selling systems. Multi-modal approach uses both textual representation of a product as well as images of given good \cite{gupte2021multimodal,wilke2021towards}. An interesting bridge between multi-modal concept and semi-surprised methodology can be found in \cite{10.1145/3397271.3401128}, where authors present self-training ensemble model called GREED.

The second family of algorithms represents the unsupervised style of learning. These models can be successfully used for product matching. Existing solutions include text-mining techniques, however, they are reported to be outperformed by supervised methods \cite{wdc}. The biggest advantage of unsupervised learning is that it can be performed on unlabelled data. The performance-to-cost-of-labeling is an inevitable trade-off that has to be faced by researchers and practitioners. This is the motivation to use a novel approach of semi-supervised algorithms as they can overcome the aforementioned trade-off. The concept of constrained clustering \cite{boosted:2018,deepclustering:2019,bair2013semi} is a great example of such an algorithm and it will be described further in the paper.

\subsection{Transforming textual data into numerical features}
Correct representation of data is a key factor in modeling. Textual data whilst it is comprehensible for humans cannot be understood the same way by machines. In order to overcome this obstacle, textual data has to be transformed into its numerical counterpart. There exist a variety of methods for achieving the aforementioned goal. These methods include a bag of words algorithm, TF-IDF %(Term Frequency-Inverse Document Frequency)
vectorization, and generating word embeddings. Another simple yet powerful concept revolves around similarity or distance measuring, proving to be successful in product matching \cite{similarity}. Examples of popular distance metrics are:
\begin{itemize}
    \item Levenshtein distance and its Damerau-Levenshtein extension \cite{levenshtein} - these are calculated based on the number of edits one has to make in order to transform one string into another. They are found in a wide spectrum of applications, inter alia, in spell checking and fuzzy string matching.
    \item Jaccard distance \cite{jaccard} - represents a token-level distance that compares sets of tokens present in both strings.
    \item Euclidean (L2) and cosine distances - they operate on word embeddings, which are  vectors generated to map words into n-dimensional space. Such transformation allows calculating of intuitive Euclidean distance as well as magnitude-independent cosine distance.
\end{itemize}

The content of features derived from the textual data can be various and is limited only by researchers' ingenuity.
%One could use NER (Named-Entity Recognition) \cite{ner} in order to tag a company name which is the manufacturer of a given product. Another approach could be based on focusing on extracting information about technical parameters of sold goods.
The selection of features is critical as attributes describing the data, regardless of undertaken methodology, have a direct impact on the performance of the model following the \emph{garbage in = garbage out} principle.

\subsection{Clustering}
Clustering is a classic example of an unsupervised learning method. It can be implemented by the standard k-means algorithm, which generates groupings based on the distances between data points and computed centroids (centers of clusters). It reassigns the cluster numbers and adjusts centroids until stopping criteria are met. Limitations of this approach do not negatively influence our solution of product matching: a number of clusters is known a priori and is equal to 2 (matching and distinct products within a pair) and all features lay within the same range of values.

\begin{figure}[htbp]
\centerline{\includegraphics{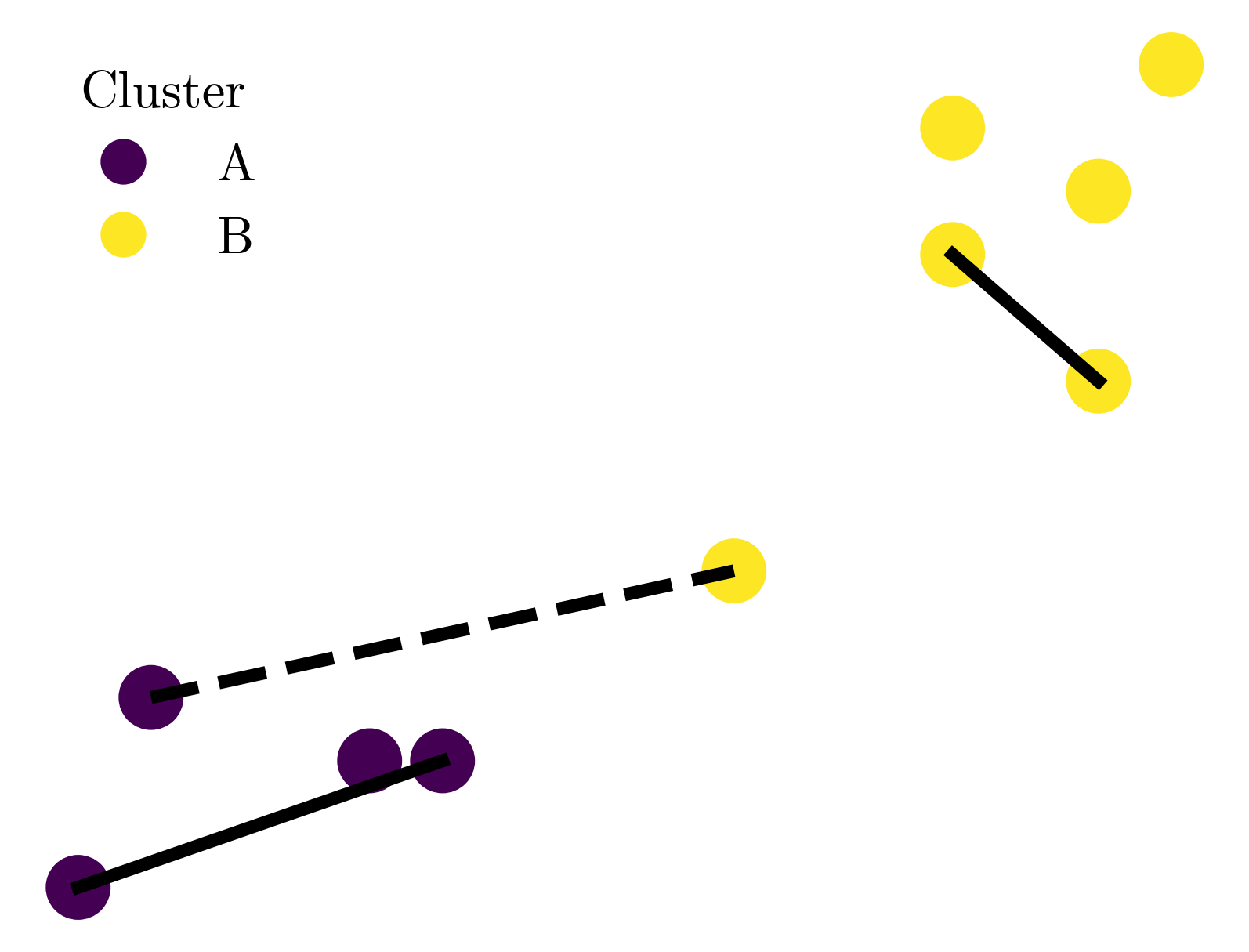}}
\caption{Example of Must Link (solid line) and Can't Link (dashed line) constraints}
\label{fig:link}
\end{figure}

An extension of the clustering algorithm -- constrained clustering -- allows feeding links describing relationships between data in such a way that they are included in the cluster assigning process. There are two types of these links:
\begin{enumerate}
    \item Must Link (ML) - pairing samples belonging to the same cluster,
    \item Can't Link (CL) - defining points that should not be grouped together in one cluster.
\end{enumerate}
Figure \ref{fig:link} presents the impact of feeding constraints into the algorithm. Note that the data point in the middle was assigned to cluster B, due to the presence of the Can't Link constraint between it and a member of a well-defined cluster A.

There are many implementations of the k-means algorithm that adopt constraints. They include COP k-means \cite{cop}, MPC k-means method, and others that can be found in \emph{conclust} package in R \cite{conclust}.

A more advanced proposal in the field of constrained clustering involves deep learning algorithms \cite{guoidec,newdeep,deepclustering:2019,dec}. Deep clustering solutions utilize autoencoders that learn the data representation themselves. The biggest contribution of DEC (Deep Embedded Clustering), and its improved version called IDEC, is the introduction of clustering loss. It is used during the training process to optimize the objective of assigning clusters. IDEC, as an extension, brings in additional losses that aim to include more advanced constraints in the learning phase. Those losses include instance difficulty loss, triplet loss, and global size constraint loss. Results obtained on benchmark datasets (MNIST, Fashion MNIST, Reuters) were reported to outperform other known methods \cite{deepclustering:2019}.

\subsection{Evaluation metrics}
Appropriate measurement of model performance is as important as any other part of the machine learning pipeline. Despite accuracy being the most intuitive way of assessing the performance it often can be deceiving. It happens when the tackled problem is defined on an imbalanced dataset. It is the case in product matching where the majority of relations between products from different feeds are of "no-match" type. In such a scenario, other metrics have to be used in order to fully describe model's ability of correct classification.

One of these metrics is the F Score often referred to as the F1 Score \cite{f1}. 
It builds upon the concepts of precision and recall which are derived from the confusion matrix. It is expressed as follows:
%\begin{equation}
%    \begin{array}{cc}
%        precision = \frac{TP}{TP + FP} \quad & \quad recall = \frac{TP}{TP + FN}\\
%    \end{array}
%\end{equation}

\begin{equation}
    precision = \frac{TP}{TP + FP}
\end{equation}

\begin{equation}
    recall = \frac{TP}{TP + FN}
\end{equation}

\begin{equation}
%\begin{split}
    F\ Score = 2 * \frac{precision * recall}{precision + recall} = \frac{TP}{TP + \frac{1}{2} * (FP + FN)},
%    \end{split}
\end{equation}
%\begin{equation}
%    F\ Score = \frac{TP}{TP + \frac{1}{2} * (FP + FN)},
%\end{equation}
where \textit{TP} refers to True Positives, \textit{FP} to False Positives and \textit{FN} to False Negatives.

Precision and recall describe quantitatively how good the model is in capturing relevant signals and how much of the captured signal is relevant. In other words, precision reflects the fraction of True Positives among all samples labeled as positive. On the other hand, recall represents how many samples were positively classified compared to all positive records.

Another metric used for evaluation in this study is strictly intended for clustering problems. Rand Index describes the similarity between two clusters \cite{rand}. It is calculated as:
\begin{equation}
    RI = \frac{number\ of\ agreeing\ pairs}{number\ of\ all\ pairs}.
\end{equation}
It checks if pairs of samples are classified the same way as in the true labels. The value of RI ranges between 0 and 1, with 1 representing entirely matching labels.

\section{Proposed Algorithm}
Our approach is based on the novel philosophy that product matching could be treated as a semi-supervised task with the knowledge about product \textit{matchings/not matchings} incorporated into the clustering constraints. Improved Deep Embedded Clustering (IDEC) is used as the constrained clustering engine. 

The solution proposed in this paper will be studied in the context of the standard dataset devoted to the task of product matching. Skroutz dataset \cite{kaggledataset} contains information sourced from online shopping platforms. For this particular research, we used the "Compact Cameras" subcategory of available product classes. We decided to use only one category of items because in such a setting the problem of distinguishing the same entities is less straightforward than comparing different groups of products. Each category uses words specific to the domain, hence intragroup data should be even better in investigating the robustness of the proposed approach. 
On the other hand, working with one category of products resembles more the routine of small retailers, as they are often specialized in given group of products and do not want to compare their goods with all available others.

In a given dataset, a single entity is described with the following fields: title, product category, and an ID used to identify same products. Products defined in this manner are paired together via cross-join and labeled if they represent the same physical item or not. The target variable was assembled based on the equality of the aforementioned IDs. Data was sampled to be imbalanced - only 25\% of 20000 generated pairs is marked as matching goods.

The textual data in order to be comprehensible to the algorithms have to be changed into numerical vectors. Features engineered for the clustering task are as follows:
\begin{enumerate}
    \item Fuzzy matching based - title ratio, title partial ratio, and title token set ratio \cite{rao2018partial}. Generated features utilize the Levenshtein distance algorithm. 
    \item Distance metric - Jaccard distance which measures dissimilarity between two sets. It is expressed as a ratio of intersection to the union of sets of all tokens found in two product titles.
    \begin{equation}
        Jaccard(P_1,P_2)=\frac{|P_1\cap P_2|}{|P_1\cup P_2|},
    \end{equation}
    where P\textsubscript{1} and P\textsubscript{2} correspond to sets of tokens found within the titles. Word tokenizer divides the string describing the product into tokens. 
    %It is done with respect to space which serves as a delimiter.
    \item Comparison of numbers found in the product titles - calculated in a similar fashion as Jaccard distance while taking only numbers into consideration. This feature proves to be useful due to its ability to compare product properties, model numbers, and technical details, which often are numerical.
\end{enumerate}
As a result, a pair of goods is represented by the numerical vector of 5 elements.

Our approach in general defines the product matching problem as a classification task. Despite the unsupervised nature of clustering algorithms, knowing true labels does not disrupt the algorithm's work, whilst allowing us to evaluate the performance of unsupervised and semi-supervised learning approaches. The task of matching products is then reformulated as an exercise of deciding if generated pair of products refers to the same entity or not. Such an approach cancels out one of the biggest shortcomings of clustering algorithms - the requirement for users to know the number of clusters prior to the calculations. In the case of the problem being solved in this paper, the number of clusters is equal to 2.

%Once the raw data were merged into a dataset of paired product titles it was possible to calculate the features. Attributes used by the k-means algorithm (which is used as an unsupervised benchmark) should be contained in the same range of values in order to minimize the risk of shifting results toward variables of the highest magnitude. Features engineered by us fit in range \textless0,1\textgreater. Standardization of attributes is a good practice given that clustering algorithms use distance-based measurements. For which reason the data was standardized.

\section{Experimental Settings and Results}

In experiments run for this research, we tested the performance of IDEC algorithm under diverging sets of constraints applied to calculations. We examined the impact of increasing the amount of Must Link and Can't Link constraints separately. The third experiment tested the influence of balance between 0-0 and 1-1 pairs in Must Link constraints. We changed only one parameter at a time in order to draw conclusions about effects directly associated with the altered setting. We ran the calculations 10 times for each set of parameters while keeping the same set of constrained pairs for those runs. 
Further tests included comparison with other available methods, performance on different datasets and measuring the impact of various data distributions.
IDEC was run with following parameters: batch size = 256, learning rate = 0.001, activation = ReLU, input dimension = 5, encoder dimensions = [200, 400, 800], decoder dimensions = [800, 400, 200], number of epochs = 20, ML an CL penalty = 1.

We split the dataset into train and test subsets. The fraction of matching pairs was preserved in both sets. Training data included 13400 samples, whereas the test set had 6600 records.

%\begin{table*}
%\centering
%\caption{IDEC hyper-parameters.}
%\begin{tabular}{|c|c|}
%	\hline 
%	\textbf{Parameter} & \textbf{Value} \\ 
%	\hline 
 %   batch size & 256 \\ 
  %  \hline
  %  learning rate & 0.001 \\ 
%    \hline 
 %   activation & ReLU \\ 
	%\hline
%	input dimension & 5 \\ 
%	\hline
%	encoder dimensions & 200 - 400 - 800 \\ 
%	\hline
%	decoder dimensions & 800 - 400 - 200 \\ 
%	\hline
%	number of epochs & 20 \\ 
%	\hline
%	ML and CL penalty & 1 \\ 
%	\hline
%\end{tabular}
%	\label{tab:idecparams}
%\end{table*}

\subsection{Constraints impact}
Using constraints in pair with clustering introduces several new parameters that can have an impact on the algorithm's performance. In this research we analyzed the significance of:
\begin{enumerate}
    \item Varying the number of Must Link Constraints - percentage numbers are presented with regards to the amount of matching pairs in the training dataset.
    \item Changing the amount of Can't Link Constraints - also references the number of "Ones" present in training. Records are sampled without replacement, hence a number of Can't Link constraints cannot exceed the number of Ones.
    \item Modifying the fraction of 1-1 pairs in Must Link Constraints - given that '1' means that products represent the same good and '0' denotes a lack of match within a pair there is a possibility of defining a Must Link pair as a relation of 0-0 or 1-1. This parameter changes the balance between pairs of both types.
\end{enumerate}
Preceding variables were changed in given ranges (where numbers in brackets represent actual number of constraints):
\begin{itemize}
    \item Must Link Constraints: 5\% (167), 10\% (335), 15\% (502), 20\% (670), 25\% (837), 30\% (1005), 40\% (1340), 50\% (1675), 60\% (2010), whilst the amount of CL was set to 10\% and the fraction of 1-1 pairs to 100\%;
    \item Can't Link Constraints: 5\%, 10\%, 15\%, 20\%, 25\%, 30\%, 40\%, 50\%, 60\%, 70\% (2345), 80\% (2680), 90\% (3015), whilst the amount of ML was set to 50\% and the fraction of 1-1 pairs to 100\%;
    \item Shifting the fraction of 1-1 pairs: from 0\% to 100\% with the interval of 10 whilst ML and CL were set to 50\% and 20\% correspondingly.
\end{itemize}

\begin{figure}[htbp]
	\centering
	\includegraphics{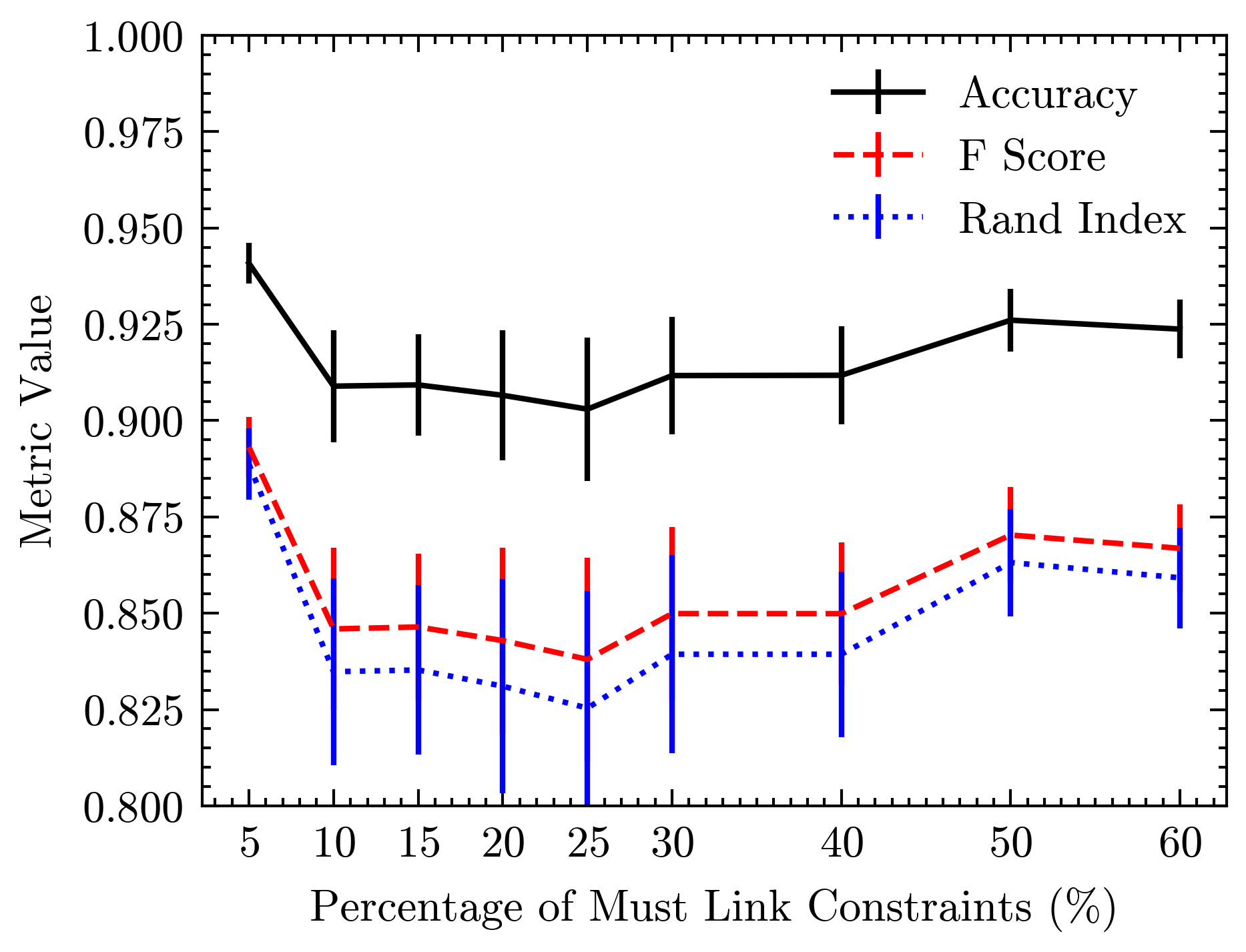}
	\caption{Impact of increasing the amount of Must Link Constraints}
	\label{fig:ml}
\end{figure}

\begin{figure}[htbp]
	\centering
	\includegraphics{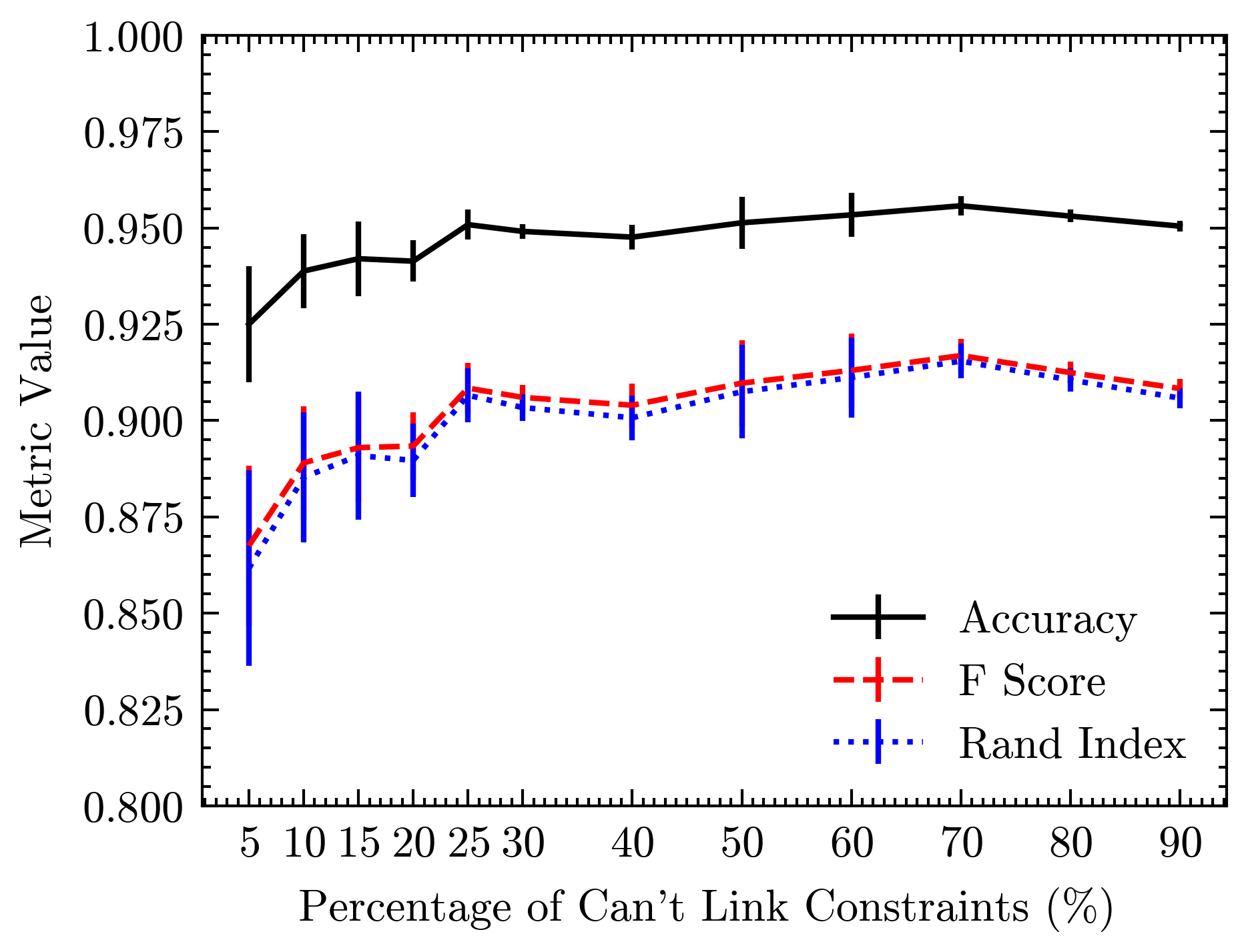}
	\caption{Impact of increasing the amount of Can't Link Constraints}
	\label{fig:cl}
\end{figure}

The effect that the amount of Must Link constraints has on the performance of clustering can be seen in Figure \ref{fig:ml}. Surprisingly, an increase in the amount of information about 1-1 pairs does not generate more accurate results. Furthermore, adding more constraints led to substantial elongation in the run time of the network training process.

Contrary to the ML constraints, expanding the set of Can't Link information about the pairs contributes to achieving better results. Figure \ref{fig:cl} depicts the relationship between the metric values and the percentage of CL constraints. The lowest F Score of 0.841 was be observed for 5\% of constraints, whereas the highest value of 0.908 is associated with run having 70\% of them. This analysis proves that sometimes less is better.
Intuition would suggest the best performance at 100\%, whereas obtained numbers show that higher amounts of information can decrease the quality of results, which is a perfect example of over-fitting. It is also important to highlight that Can't Link constraints are easier to obtain in real-world scenarios - as they could be generated automatically, without costly surveys of retail experts. 

\begin{figure}[htbp]
	\centering
	\includegraphics{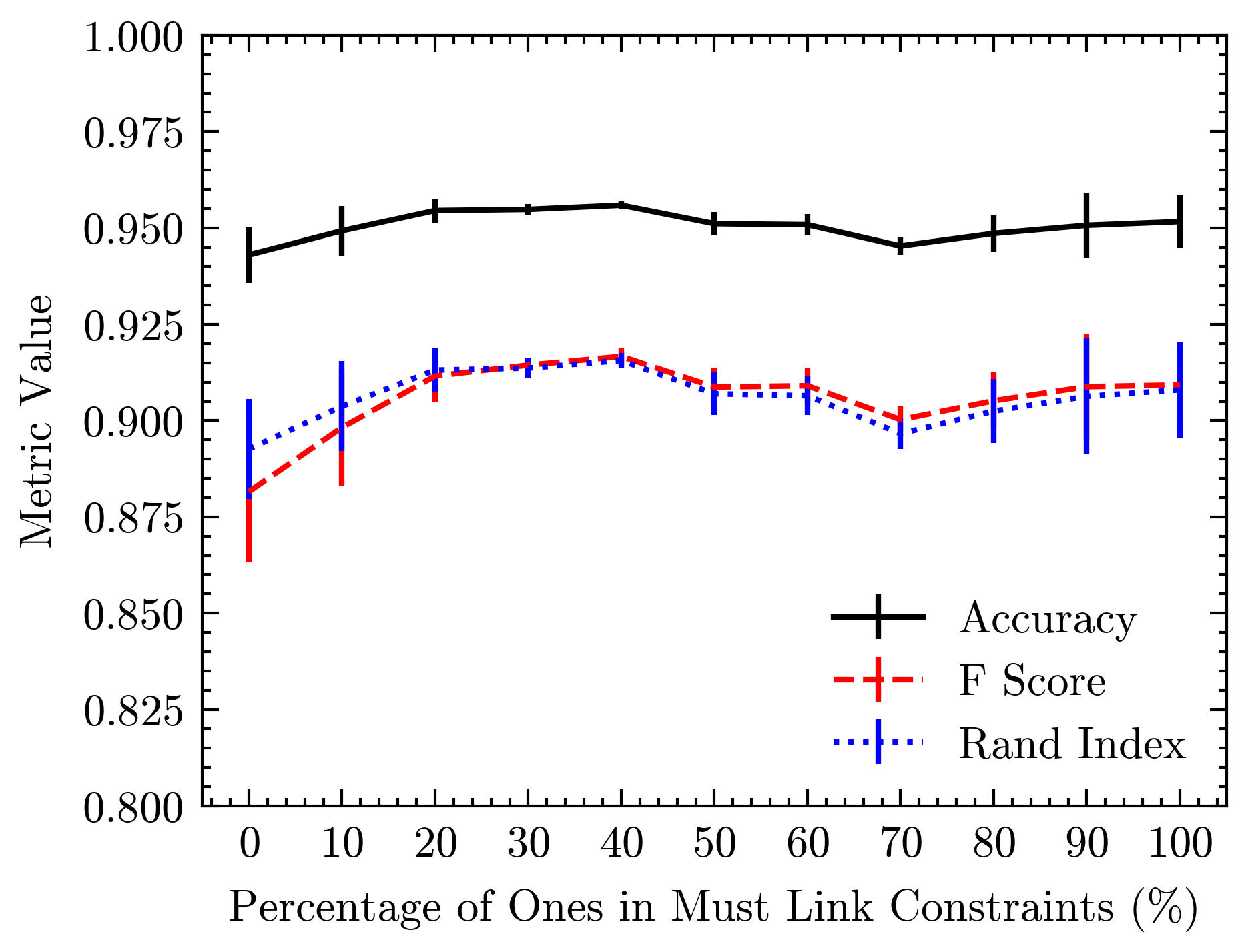}
	\caption{Impact of increasing the amount of 1-1 pairs in Must Link Constraints}
	\label{fig:fr}
\end{figure}

Figure \ref{fig:fr} shows where the perfect balance between types of Must Link Constraints lies. Observation of F Score curiously shows that the best results are achieved when there are 20\% of 1-1 (actually matching products constraints) pairs and 80\% of 0-0 pairs. We suspect that given the imbalanced nature of the data, increasing the signal for True Positives is treated by the model as forcing the outliers to be grouped together. Then instead of guidance on how to cluster points, the model gets explicit instructions resulting in over-fitting and therefore reducing the overall performance. This hypothesis is also supported by the results of the first experiment (Figure \ref{fig:ml}) where given a constant balance between types of pairs, a higher amount of constraints (simply increasing volume of 1-1 pairs) presented to the model did not improve its quality.

\subsection{Comparison with other methods}
The experiment designed in this paper aimed to investigate the advantage of semi-supervised algorithms in product matching over the traditional unsupervised and supervised learning represented by the k-means and XGBoost methods, respectively. In addition, more modern method was used to assess the product pairs. DeepMatcher algorithm \cite{deepmatcher} implements 3 ways of solving entity matching problems: \textit{SIF (Smooth Inverse Frequency)} which is a simple, aggregate model; \textit{RNN} that takes sequence into consideration and \textit{Attention} aware of both sequences and data similarity. 

F Score was maximized in order to select the best threshold for DeepMatcher classification. Results selected in that manner were then compared with best runs of IDEC, k-means and XGBoost with respect to the set of constraints discovered in previous experiment. 

\begin{table*}
\centering
\caption{Comparison of k-means, IDEC, XGBoost and DeepMatcher algorithms.}
\begin{tabular}{|c|c|c|c|c|}
	\hline 
	\textbf{Algorithm} & \textbf{Experiment} & \textbf{Accuracy} & \textbf{F Score} & \textbf{Rand Index} \\ 
	\hline 
	k-means & & 0.924 $\pm$ 0.0001 & 0.848 $\pm$ 0.0004 & 0.859 $\pm$ 0.0002 \\ 
	\hline 
	IDEC-ML5-CL10-F100 & Must Link & 0.941 $\pm$ 0.0052 & 0.893 $\pm$ 0.0078 & 0.889 $\pm$ 0.0093 \\ 
	\hline
	IDEC-ML50-CL70-F100 & Can't Link & 0.956 $\pm$ 0.0025 & 0.917 $\pm$ 0.0044 & 0.915 $\pm$ 0.0045 \\
	\hline
	IDEC-ML50-CL20-F40 & Fraction 1-1 & 0.956 $\pm$ 0.0011 & 0.917 $\pm$ 0.0023  & 0.916 $\pm$ 0.0020 \\
	\hline
	XGBoost-ML5-CL10-F100 & Must Link & 0.865 $\pm$ 0.0213 & 0.782 $\pm$ 0.0290 & 0.767 $\pm$ 0.0307 \\ 
	\hline
	XGBoost-ML50-CL70-F100 & Can't Link & 0.871 $\pm$ 0.0234 & 0.789 $\pm$ 0.0339 & 0.777 $\pm$ 0.0352 \\
	\hline
	XGBoost-ML50-CL20-F40 & Fraction 1-1 & 0.910 $\pm$ 0.0242 & 0.837 $\pm$ 0.0407  & 0.837 $\pm$ 0.0339 \\
	\hline
    DeepMatcher-attention & & 0.934 $\pm$ 0.0076 & 0.838 $\pm$ 0.0158 & 0.877 $\pm$ 0.0133 \\
    \hline
    DeepMatcher-rnn & & 0.894 $\pm$ 0.0063 & 0.791 $\pm$ 0.0089 & 0.811 $\pm$ 0.0099 \\
    \hline
    DeepMatcher-sif & & 0.859 $\pm$ 0.0197 & 0.643 $\pm$ 0.0314 & 0.758 $\pm$ 0.0283 \\
    \hline
\end{tabular}
	\label{tab:results}
\end{table*}

Table \ref{tab:results} presents the results of the best runs in each category of diverging parameters. The best run was picked with respect to the F Score as it is the most descriptive of the algorithm's performance given the task being solved. It reports the average and standard deviation calculated over all runs with particular settings. Results demonstrate that enriching data with some additional information, namely constraints, leads to higher quality results. An increase of 0.07 in the F Score measurements was observed, compared to the k-means algorithm. Despite this being a relatively small improvement it can lead to a substantially significant gain in the task of finding True Positives - pairs of products that match. This effect is especially desired having in mind the real-world application of the presented framework. The amount of real data that can be analyzed with this methodology far exceeds the sample of 20000 records used in this study. In big-data problems, even the smallest upgrade goes a long way.

Analysis of XGBoost results and its mirrored IDEC runs show that Deep Clustering performs better in all of the measured scenarios. It is worth mentioning that Deep Clustering approach outperformed the k-means algorithm in every reported metric. DeepMatcher, regardless of the used model, performs worse than proposed algorithm with respect to all evaluation metrics.

To further compare k-means and IDEC algorithms Table \ref{tab:misclass} presents examples of product pairs that were impossible for the simple algorithm to classify correctly. Analysis of these results allows drawing the conclusion that IDEC is more versatile and robust due to its ability to operate regardless of the length of compared strings, presence of typos, or distinctive ways of presenting product parameters.

\begin{table*}[htbp]
\centering
\caption{Examples of pairs misclassified by k-Means while being correctly classified by IDEC.}
%\scalebox{0.98}{
%\begin{tabular}{|P{0.5cm}|P{5cm}|P{5cm}|P{3cm}|}
\begin{tabular}{|c|c|c|}
	\hline 
	\textbf{Product 1} & \textbf{Product 2} & \textbf{Type}  \\ 
	\hline 
	%canon powershot g9x mark ii black pliromi ke se eos 36 dosis & canon powershot g9 x mark ii 20.1mp 1 cmos 5472 x 3648pixels mavros mavro & TP\\ 
	%\hline
	panasonic lumix bridge camera dc fz82 eu ka269712 & panasonic lumix fz82 black eos 24 dosis i eos 60 dosis choris karta & TP\\ 
	\hline 
	nikon coolpix a100 purple thiki nikon doro vna974e1 & compact fotografiki nikon coolpix a100 purle se 3 atokes dosis & TP\\ 
	\hline
	aquapix w1024 b 10017 adiavrochi kamera mavri 10 mp & easypix aquapix w1024 splash red & TP\\
	\hline 
	olympus tough tg 5 digital camera & olympus tg 5 red eos 24 dosis i eos 60 dosis choris karta & TP \\
	\hline
	sony dsc hx60 & sony dsc rx10 iii & TN\\
	\hline
	%panasonic lumix dmc ft30 blue & panasonic lumix dmc fz 82 & TN\\
	%\hline
	canon powershot sx730 hs silver & canon powershot sx620 hs red 1073c003 & TN\\
	\hline
	fotografiki michani nikon coolpix a100 red & fotografiki michani olympus tg 5 red & TN\\
	\hline
	cybershot dsc rx100m3 & sony cybershot dsc rx10 m2 se 12 atokes dosis & TN\\
	\hline

\end{tabular}
%}
	\label{tab:misclass}
\end{table*}

\subsection{Other datasets}
Additionally, to fully examine quality of the proposed algorithm it was run on other data. Except for "Compact Cameras" category from Skroutz dataset the mixture category of other camera related categories was generated. This dataset called \textit{cameras all} includes Analog Cameras, Mirrorless Cameras and Digital Single Lens Reflex Cameras. As for benchmarking standards a WDC (Web Data Commons) dataset was used for more calculations. Data about cameras (\textit{wdc cameras}) from Version 2.0 of Large-Scale Product Matching Dataset \cite{wdc_data} was employed to further check robustness of the semi-supervised approach. IDEC algorithm used constraints combination ML5-CL10-F100.

%\begin{figure}[htbp]
%	\centering
%	\includegraphics{datasets.png}
%	\caption{IDEC performance on various datasets}
%	\label{fig:datasets}
%\end{figure}

\begin{table}[htbp]
    \centering
        \caption{IDEC performance on various datasets.}
    \begin{tabular}{|c|c|c|c|}
        \hline
         \textbf{Dataset} & \textbf{Accuracy} & \textbf{F Score} & \textbf{Rand Index} \\
         \hline
         cameras & 0.941 $\pm$ 0.0052 & 0.893 $\pm$ 0.0078 & 0.889 $\pm$ 0.0093 \\
         \hline
         cameras all & 0.899 $\pm$ 0.0296 & 0.815 $\pm$ 0.0375 & 0.820 $\pm$ 0.0452 \\
         \hline
         wdc cameras & 0.789 $\pm$ 0.0284 & 0.665 $\pm$ 0.0505 & 0.668 $\pm$ 0.0313 \\
         \hline
    \end{tabular}
    \label{tab:datasets}
\end{table}

Table \ref{tab:datasets} shows consistent, high performance achieved on diverse data sources. A drop in metric values for \textit{wdc cameras} data might be caused by the fact that given dataset mixes product feeds from many pages and multiple languages. In that fashion some pairs are cross lingual, which might be hard for fuzzy features to reflect in the training data.

\subsection{Class distribution impact}
\begin{figure}[htbp]
	\centering
	\includegraphics{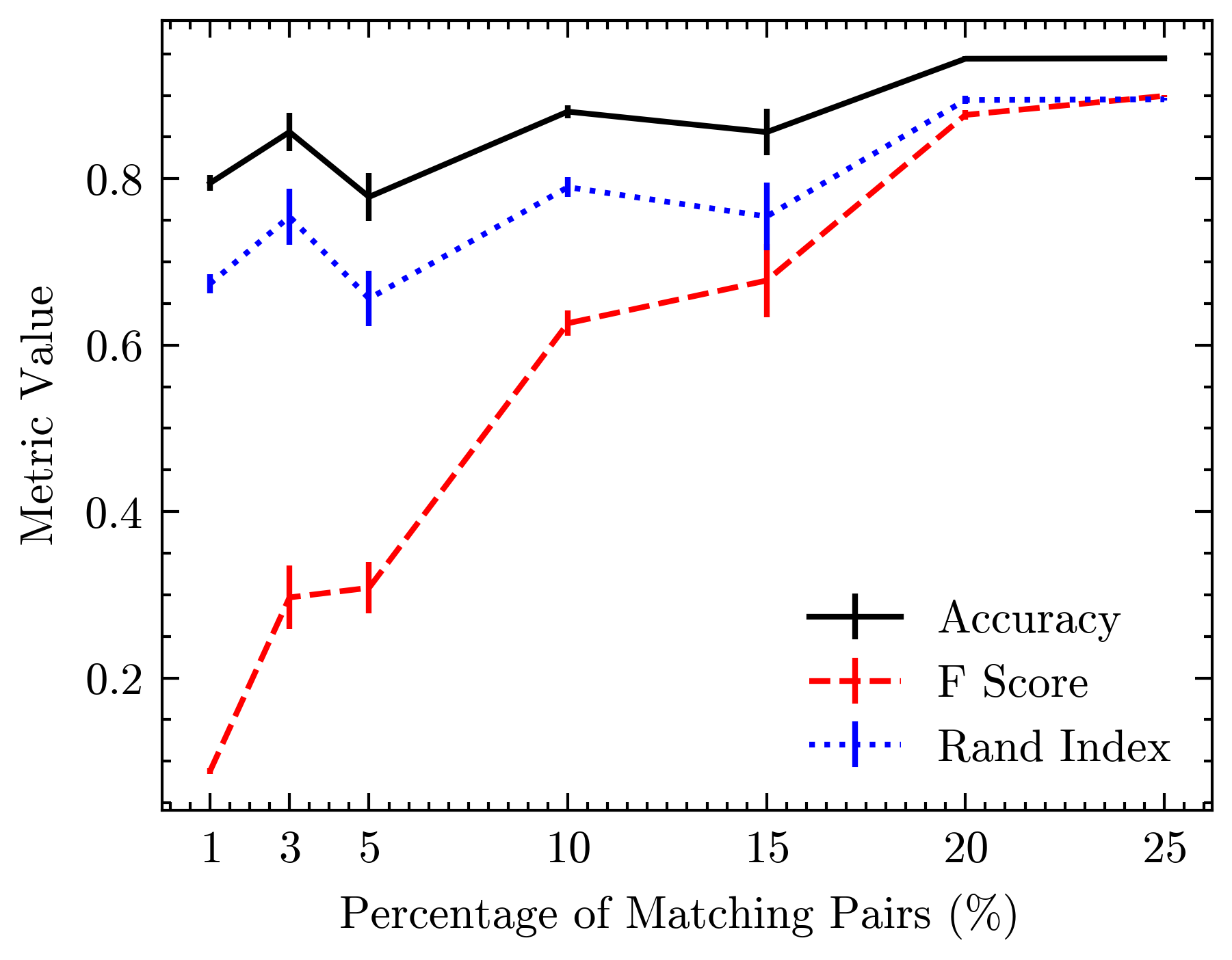}
	\caption{Impact of increasing the amount of matching pairs in the dataset}
	\label{fig:dist}
\end{figure}

Another quality assessment test touches the subject of algorithm's performance at varying data distributions. For this experiment datasets with increasing percentage of matching pairs (ones) were synthetically generated from the \textit{cameras} data. Percentages used for these datasets were as follows: 1, 3, 5, 10, 15, 20, 25. IDEC algorithm was run with the same constraints setup as in previous experiment (ML5-CL10-F100). Figure \ref{fig:dist} presents increasing trend for all of reported metrics. The threshold of 10\% brings substantial increase of F Score value, if doubled all metrics are performing over the value of 0.8.

It is worth mentioning that even WDC Gold Standard use datasets whose percentage of positive pairs range from 37\% to 48\% in the training data, for classes of shoes and computers respectively. It is far from the tested and real life scenarios.

The implementation of the tested algorithms along with the full results of this study can be found in the repository [{URL} undisclosed for peer-review].

\section{Conclusion}
The given paper presents a framework that can be utilized for the problem of product matching. The nature of the data and given task requires the solution to be functional with as little information as is possible. Online shopping websites often do not bear consistent information across all offerings not to mention cross-platform discrepancies. The proposed solution uses only titles of the products and derives simple text-based features which, unlike generating word embeddings, do not require high computational power.

Results obtained with the Deep Clustering approach outperform standard k-means algorithm, XGBoost method as well as deep learning based solutions. These results demonstrate that enriching data with constraints, transforming the problem from the unsupervised to the semi-supervised domain, leads to a positive outcome. An increase in \makebox{F Score} metric of 0.07 can have a high impact on real-world applications of such product matching concepts.
More advanced tests proved presented solution to be robust and high performing on diverse datasets as well as under various class distributions.

There exist multiple ways of improving this research and extending it to examine more possibilities for boosting the model's results. We tested IDEC performance while changing only one parameter at a time, but this could be expanded to diverging at least two attributes concurrently. Another path for improvement could utilize alternative existing constrained clustering algorithms such as COP k-means or the MPC version of it. Adding more features, for example related to the spread of prices of products, can be a valuable extension of the modeling process as well.

It has to be pointed out that within the subject of product matching of online offerings datasets are countless. In the era of widely available web scrapers, there is a possibility of gathering various features at the very source of the data - the on-line selling platforms. Solutions, as well as the data volumes, are only limited by the computational costs of running algorithms and storing records. For these reasons, product matching is an important task worth further research and development.

%\begin{table}[htbp]
%\caption{Table Type Styles}
%\begin{center}
%\begin{tabular}{|c|c|c|c|}
%\hline
%\textbf{Table}&\multicolumn{3}{|c|}{\textbf{Table Column Head}} \\
%\cline{2-4} 
%\textbf{Head} & \textbf{\textit{Table column subhead}}& \textbf{\textit{Subhead}}& %\textbf{\textit{Subhead}} \\
%\hline
%copy& More table copy$^{\mathrm{a}}$& &  \\
%\hline
%\multicolumn{4}{l}{$^{\mathrm{a}}$Sample of a Table footnote.}
%\end{tabular}
%\label{tab1}
%\end{center}
%\end{table}

%\section*{Acknowledgment}
%Placeholder for acknowledgments?

\bibliographystyle{ieeetr}
\bibliography{mybib}
\end{document}